# ON THE CRITICAL IONIZATION VELOCITY EFFECT IN INTERSTELLAR SPACE AND POSSIBLE DETECTION OF RELATED CONTINUUM EMISSION

Gerrit L. Verschuur

*Physics Department, University of Memphis*

*Memphis, TN 38152*

gverschr@memphis.edu

   *Abstract*—Interstellar neutral hydrogen (HI) emission spectra manifest several families of linewidths whose numerical values (34, 13 & 6 km/s) appear to be related to the critical ionization velocities (CIVs) of the most abundant interstellar atomic species. Extended new analysis of HI emission profiles shows that the 34 km/s wide component, probably corresponding to the CIV for helium, is pervasive. The 34 km/s wide linewidth family is found in low-velocity (local) neutral hydrogen (HI) profiles as well as in the so-called high-velocity clouds. In addition, published studies of HI linewidths found in the Magellanic Stream, Very-High-Velocity Clouds, and Compact High-Velocity Clouds, all of which are believed to be intergalactic, have noted that typical values are of the same order. If the critical ionization velocity effect does play a role in interstellar space it may be expected to produce locally enhanced electron densities where rapidly moving neutral gas masses interact with surrounding plasma. Evidence is presented that suggests that this phenomenon is occurring in interstellar space. It manifests as a spatial association between peaks in HI structure offset with respect to peaks in high-frequency radio continuum data obtained with the Wilkinson Microwave Anisotropy Probe.

I. Introduction

Interstellar neutral hydrogen (HI) gas emits a spectral line at 21-cm wavelength that is readily observed using radio astronomical techniques. The observed emission profiles have a shape determined in part by thermal broadening to produce a line width measured at half maximum brightness of order 2 km/s for 80K gas, and in part by random bulk motion such as may be introduced by turbulence in an HI cloud. Superimposed on this, the spectral line will also be shifted with respect to its rest wavelength by a Doppler shift due to bulk motion along the line-of-sight. There may be other phenomena acting to influence the observed spectral line shape and this report will discuss one of those. The key point is that by analyzing the shape of 21-cm emission profiles it is possible to learn about the physics of the gas involved. The work described here concerns the shapes of the 21-cm emission profiles in directions well away from the plane of the Milky Way galaxy where the typical lines of sight encompass HI within a few hundred parsecs of the sun, depending on galactic latitude (the angular distance above the central plane of the Milky Way), given that the galactic disk is of order 200 pc thick with the sun approximately in the center.

In a simple model for the Galaxy, nearby gas shares a motion not very different from that of the sun, and hence there would only be a small Doppler shift of order 10 to 20 km/s due to random bulk motions known to be typical of interstellar space to produce the observed emission profiles. What is observed is far more complicated, however. Local HI with velocities around zero km/s with respect to the local standard of rest (l.s.r. - the standard defined by radio astronomers, who display HI emission profiles as a function of this velocity) usually dominates the high galactic latitude profiles but there is also a great deal of emission around –50 km/s, produced in a population of so-called intermediate velocity clouds, as well as about –125 km /s for so-called high-velocity clouds. An unambiguous estimate for the distance to these HI features has not yet been agreed upon although there is little doubt that they are galactic.

Verschuur & Peratt [1] have described the Gaussian decomposition of interstellar HI emission profile shapes and find that at least three families of linewidths of order 34, 13 and 6 km/s characterize the profiles. If the linewidths are taken to indicate kinetic temperature these would imply that the HI along a given line of sight manifests temperatures of order 24,000, 3,500 & 750 K. However, at the former value the hydrogen atoms should be ionized and hence not observable at 21-cm wavelength. Recently published results of a more extensive study [2] confirm the existence of these linewidth families, which are most clearly seen in directions where the overall HI profiles have relatively low column densities, $< 10^{20}$ cm $^{-2}$.

Here we report that the 34 km/s wide components appear to be pervasive in an area of sky in the northern galactic hemisphere studied in detail, an area of 4,800 square degrees where the HI low-velocity (v≤30 km/s) distribution is severely disturbed and where intermediate velocity (-100<v<-30 km/s) and high-velocity (<-100 km/s) HI features are observed. It is beyond this report to discuss the lack of the 34 km/s wide component in the southern galactic hemisphere except to note that in those direction there is little evidence for an underlying velocity disturbance [3].

It has been suggested [1], [4] that the linewidth families are related to a plasma phenomenon known as the critical ionization velocity (CIV) effect based on a numerical similarity between families of CIV values and HI linewidths. The CIV is defined as that velocity at which a neutral particle traveling into plasma permeated by a magnetic field becomes ionized when its kinetic energy is equal to its ionization potential [5]. The numerical values of the HI linewidth families are similar to the values of Bands of CIVs for the most abundant atomic species; 34 km/s for helium, 13 km/s for C, N & O, and about 6 km/s for heavy metals such as Si, Al, and Fe [1], [4].

If the CIV phenomenon plays a role in interstellar space, we may expect that at any surface between a rapidly moving, neutral gas mass and the surrounding plasma (also known as the intercloud medium), enhanced electron densities should be produced. These electrons, in turn, might be capable of generating weak radio frequency continuum (broad band) emission. Whether or not the observed spectrum of such radiation is thermal or non-thermal depends on local conditions in the interface region. For example, in the simplest case weak thermal emission might be expected, but if plasma instabilities were to be triggered by the interaction between the moving gas mass and the ambient plasma, particle acceleration might occur [6] to generate non-thermal emission. Here we report on the discovery of an association between HI structure and high frequency continuum radiation but the spectrum of the latter has yet to be determined.

For this study we used the side-lobe correct neutral hydrogen (HI) emission profile data from the Leiden-Dwingeloo All-Sky HI Survey data [7] as well as some observations from the Leiden-Argentina-Bonn (LAB) All-Sky HI Survey [8].

## II. Linewidth Histograms

Figures 1a and 1b reproduce the Gaussian component linewidth histograms for a sample of profiles at high positive and negative galactic latitudes, mostly between galactic latitudes 70° and 80° from [2]. Fig. 1a is the histogram of the linewidth of components for northern galactic hemisphere data with center velocities between -7 and -30 km/s that show the presence of the 34 km/s wide component most clearly. In the case of the southern hemisphere data (Fig. 1b), virtually all the low-velocity (LV) components are found with center velocities between - 9 and +3 km/s whereas in the northern hemisphere the LV distribution extends from -30 to +10 km/s. Superimposed on both figures is a curve that represents the histogram of linewidths for the Compact High-Velocity Clouds (CHVCs) [9]. These are hypothesized to be extragalactic objects in the Local Group [10], [11]. The relevant linewidth data for the CHVCs are taken from Table 3-2 of reference [10].

In Fig 1a there is a virtually perfect agreement between the two histograms over the linewidth range from 30 to 45 km/s and this coincidence piqued our curiosity. Why do these histograms agree so closely, given that the data include only local HI at high galactic latitudes [2] while the CHVCs may be extragalactic? It is striking that the data in Fig. 1a also agree on the presence of a wing in the linewidth distribution to around 45 km/s.

## III. Comparison with Other Data

The 34 km/s wide family of HI components is found in high-latitude, low-velocity HI [1], [2], in Very High-Velocity Clouds [12], the Magellanic Stream [13], [14] and Compact High-Velocity Clouds [9]. These latter three categories of HI structures are believed to be located in relatively nearby (≤50 kpc distant) intergalactic space.

The median linewidth for the CHVCs is 35 km/s [9]. Analysis of our data (Figure 1) gives median values for the two data sets of 34 km/s and 33 km/s. Furthermore, our observations of high-velocity HI profiles [3] that could be fit by a single Gaussian component produce a median linewidth of 34.0 km/s. The typical linewidth of a sample of Very-High-Velocity Clouds in Cetus is 30 km/s [12]. Our analysis of a sample of profiles toward the tail of the Magellanic Stream produced a median linewidth of 36.3 km/s, which compares with published values for HI in the Stream of 34 km/s [13], 32 km/s [14] and 32 km/s [15]. The median value for the corresponding component is 31.0 km/s for low velocity HI [1]. Verschuur & Schmelz [16] first reported on the existence of this family of linewidths and their data indicate a median value of 36.3 km/s. A median linewidths of 36 km/s has also been reported for several other categories of HI profiles including high-velocity "clouds" [9].

These various results all point to the pervasive presence of a family of linewidths in HI emission profiles of order 34 km/s. Other than our arguments, it is not obvious that the other authors considered the significance of HI profiles with the broad linewidths. In general it has been regarded as evidence for the existence of a warm intercloud medium without ever explaining why HI would still be in the neutral state at the temperature of about 24,000 K implied by this linewidth.

IV. Observing the 34 km/s Line in a Direction of Low Column Density

As part of what was initially an unrelated project, the HI distribution in an area of sky shown in Figure 2 was mapped and a Gaussian analysis performed at 1° intervals in galactic longitude from l=135° to 148° and 0.°5 intervals of galactic latitude from b=60.°5 to b=63.°0. What was found is that the low velocity gas surrounding the HI peaks seen in Fig. 2 could best be fit by a single Gaussian of order 34 km/s wide compared to about 23 km/s for directions enclosed by the closed contours defining "clouds" of HI. If taken at face value this implied that the "background" HI seen around the peaks in Fig. 2 was intimately tied to the existence of those peaks, whose linear extent would be small given that their structure is of order a degree or so across. Yet the path length of the line-of-sight through the disk of the galaxy is of order 116 pc in this direction (for the galactic latitude of 62° with the sun located at the center of a galactic disk of order 200 pc thick). An order of magnitude argument notes that the angular extent of the peaks (or "clouds") seen in Fig. 2 is about 1°, which corresponds to 1 pc at a distance of 57 pc. This diameter is a small fraction of the ~100 pc path length through the galactic disk for the area in Fig. 2. If the 34 km/s wide components, which encompass all of the HI at low-velocities in this direction, are only associated with the small-scale structure see in Fig. 2 whose linear extent is of order 1 pc, it would imply that the rest of the line-of-sight is devoid of low-velocity HI. This is physically unlikely. Instead, the 34 km/s Gaussian components seen around the edges of the peak emission in Fig. 2 represents a signal that encompasses all the low-velocity HI in the line-of-sight through the galactic disk in those directions. This implies that in order to perform a meaningful Gaussian analysis of the emission profiles in the area of the peaks in Fig. 2, the underlying, quite real, 34 km/s wide component must first be identified and then subtracted, which was therefore done.

The parameters of the pervasive 34 km/s wide component were determined by examining the average profile shapes around the edges of the peak emission where the low velocity HI profile was readily fit by a single component.

The same process of first identifying and then removing an underlying 34 km/s wide component from the observed HI profiles was applied to a similar HI structure found at $l=85°$, $b=58°$ and Figure 3 shows the linewidth histograms for the low-velocity components (center velocities > -30 km/s) for these two areas combined. The signatures of the 13 and 6 km/s linewidth families [1] are prominent. Bear in mind that the 34 km/s wide component is not seen in this histogram because it was first removed as described above.

V. The Origin of 34 km/s Wide HI Emission Components

It is possible that the 34 km/s component in the studies referred to above is made up of much narrower components. Something like this has been reported in the case of the small angular scale of structure in high-velocity cloud (HVC) A1 [17]. Emission line components in HVC A1 as narrow as 3 km/s have been observed with a resolution of 1 arcsecond [18] and the center-velocity histogram of those components produces an overall distribution 20 km/s wide, which corresponds to the linewidth for that location in HVC A1 when observed with a 10 acrminute beam. Other data also exist that show that when a broad HI component in a small area of sky is observed with an aperture synthesis telescope the profile structure does indeed reveal the presence of narrower components (colder gas) and that the ensemble velocity distribution accounts for the broader feature observed with lower angular resolution [17].

If the 34 km/s wide features discussed above consisted of multiple narrow lines when viewed with sufficiently high resolution, this would avoid the concern raised in trying to account for HI with an apparent kinetic temperature of 24,000 K (implied by the linewidth of 34 km/s). However, this would not remove the question of why an ensemble of unresolved HI concentrations would, in so many instances, have a Gaussianvelocity distribution whose width is of order 34 km/s. It is important to note that the 34 km/s wide component family of linewidths was first noticed using data obtained with the Arecibo radio telescope with a beamwidth of 2 arcminutes [16]. Thus it is not regarded as likely that this broad component is made up of an ensemble of narrow lines.

Turbulence might be capable of producing an apparent linewidth of 34 km/s for an ensemble of small cells. However, it is then difficult to account for the uniformity of the linewidth data for local high latitude HI, the Magellanic Stream, VHVCs, and CHVCs. These four categories surely exist in different environments and there is no obvious a priori reason to expect turbulence to be the same in all four.

A multi-component model for interstellar clouds has enjoyed much popularity over the years. In such a model, a cold "cloud" of HI is said to be in pressure equilibrium with a warm "intercloud" medium. However, our data suggest the existence of at least 3 linewidth families. Using only the data for the background profiles in the area of Fig. 2, the HI column density for the low-velocity component along the line-of-sight is $3.2 \times 10^{19}$ cm$^{-2}$. For the path length of 116 pc estimated above, this implies an average volume density of 0.1 cm$^{-3}$. Thus, if the 34 km/s wide component corresponds to an "intercloud" medium, its pressure, nT, is of order 2,400 cm$^{-3}$ K. If the low velocity peaks in the HI map shown in Fig. 3 are then interpreted as evidence for "clouds" at a distance of 57 pc, say, with diameters of order 1°, their volume densities can readily be estimated from the Gaussian analysis. These are 4.8 cm$^{-3}$ for the 13 km/s wide lines and 3.1 cm$^{-3}$ for the 6 km/s wide lines. If the linewidths are interpreted as a simple indictor of kinetic temperature these two classes of component are at 3,700 K and 610 K and their internal pressures would then appear to be 17,760 cm$^{-3}$ K and 1,891 cm$^{-3}$ K respectively, which compares with the pressure in the background component, 2,400 cm$^{-3}$ K. Only the HI related to the family of linewidths of order 6 km/s would be likely to be in pressure equilibrium with the HI producing the 34 km/s wide emission, if a cloud model is invoked to explain the data. However, this then raises questions related to why the 13 km/s wide family of linewidths with a very diferent implied pressure would be found within these same "clouds."

## VI. Is the Critical Ionization Velocity Phenomenon Involved?

In what follows we attach significance to the numerical coincidence between the CIV of helium (34.4 km/s) and the typical linewidth of the broad background component seen in HI emission profiles (~34 km/s). However, it cannot be ruled out that molecular hydrogen with a CIV of 38.5 km/s also contributes to this component within the uncertainties in the data.

Reviews of the CIV phenomenon make it clear that it is poorly understood phenomenon [19], [20], [21]. They do agree that the process requires the generation of plasma instabilities that result in some particles being accelerated to high enough energies to trigger ionization. Given that the mechanism by which the CIV effect is believed to operate in the laboratory and in the earth's magnetosphere is not fully understood, data on the presence of the helium CIV signature in interstellar HI emission profiles may, instead, help understand the physics of the mechanism. By implication, the same process would need to be invoked to account for the other families of linewidths discussed here, those around 13 km/s, associated with the CIVs of C, N & O, and 6 km/s, associated with heavy metals such as Al, Fe, Si, Na & Ca.

While a full discussion as to why the CIV effect might influence the 21-cm interstellar HI spectral line shapes is beyond the scope of this paper, it is worth noting that the CIV of hydrogen atoms is 51 km/s and yet the signature of this CIV has not been detected in the interstellar HI profiles. There are several possible explanations for this. A component 51 km/s wide would be difficult to identify in the weak emission profiles, in particular because of the close proximity of the local, low velocity HI with respect to the intermediate velocity emission as regards the velocity spread. A search for a weak, 51 km/s wide component would require a directed research program, probably requiring new, high-resolution data. More relevant, however, is the issue of whether one would expect a component of this linewidth in the first place.

If we accept as a working hypothesis that the 6, 13 and 34 km/s families of HI component are indicative of the action of the CIV effect in interstellar space, then we are observing the consequence of coupling between neutral hydrogen atoms and the ions and electrons created by the CIV effect acting on other atomic species. However, if the CIV effect also ionizes the hydrogen where the relative velocity of the neutral gas with respect to a plasma is large enough, we would not expect to observe 21-cm radiation because, by definition, the hydrogen is then ionized. This means that the 51 km/s CIV will not be reflected in the HI emission profile shapes because there are no neutral particles available to produce the 21-cm radiation. Instead, we might expect to find H-$\alpha$ emission from the ionized hydrogen. This has been observed toward several high- and intermediate-velocity HI clouds [22]. More highly ionized gas has also been found to be associated with high-velocity clouds, which suggest that the ionization is produced at the interfaces between rapidly moving HI structures and surrounding (galactic) coronal gas [23]. Others have drawn a different conclusion, claiming that the H-$\alpha$ emission proves that the high-velocity gas is immersed in a hot medium [24]. However, we suggest that high-velocity gas could be creating the hot medium during its passage through interstellar space by the action of the CIV effect. This has already been proposed [25] to account for H-$\alpha$ emission observed toward the Magellanic Stream [26].

Another piece of circumstantial evidence that points to the possibility that the CIV effect may be occurring in interstellar space is related to the fact that the so called "intermediate velocity clouds," or IVCs, have a typical velocity of about -50 km/s with respect to the l.s.r. No one has ever successfully explained the magnitude of this velocity. When the center velocity of nearly 4,000 Gaussian components in the direction of IVCs and related high-velocity HI is plotted an average velocity for 1,263 intermediate velocity components is $-51.7 \pm 6.0$ km/s and for 794 low-velocity components is $+0.7 \pm 2.5$ km/s [3]. There appears to be a real cut-off in the HI distribution to more negative velocities in excess of $-51$ km/s at which velocity the HI would be ionized due to the CIV effect. It is also rather startling that for those directions where the IVC center velocity is shifted to slightly more negative velocities (for example to $-65$ km/s), the low velocity gas velocity center in the same direction is also shifted by about 15 km/s, or is almost totally missing.

We have found some evidence for the presence of an HI component of order 50 km/s wide, but it is related to the presence of weak positive velocity emission at northern galactic latitudes, something that is beyond the thrust of the present report because the very existence of positive velocity emission (in the range +30 to +50 km/s) is forbidden in this area of the sky (on any model of galactic structure), something that has yet to be discussed in the literature.

No other studies other than those referred to above {1], [2} and [4] have appeared that consider the possible role of the CIV in interstellar space, although it has been considered as a likely cause for H$\alpha$ emission from the Magellanic Stream [25]. Clearly, this field or research is a wide open for further study.

VII. Associations between Continuum Radio Frequency Emission and HI Structure

If the CIV effect does operate in interstellar space, regions of locally enhanced electron density will be expected at the interfaces between moving neutral gas masses and any ambient plasma. An observational test of what might be occurring in those regions where the CIV effect is hypothesized to be taking place is to search for weak radio frequency emission from newly created thermal electrons, or from electrons that may, in addition, have been accelerated to high energies by the action of plasma instabilities at the interface between the moving neutral gas mass and the plasma.

To test our hypothesis, we drew upon the all-sky survey of high-frequency radio continuum [27] based on observations obtained by the Wilkinson Microwave Anisotropy Probe (WMAP) spacecraft. Its original goal was to map the small-scale structure of the cosmic microwave background believed to have its origin in the aftermath of the Big Bang. The observations were made at 5 frequencies from 23 to 94 GHz and the WMAP team combined those data and removed the effect of known foreground emission to produce a single summary plot known as the Internal Linear Combination (ILC) map. Clearly, if the small angular scale structure observed by WMAP does originate at cosmological distances there should be no relationship to galactic HI structure. However, as will be shown below, find evidence for such a relationship has been found. This has far-reaching implications for the interpretation of the WMAP data, discussion of which is beyond the scope of this report. Instead, we treat the WMAP survey, specifically from the (first-year) ILC map, as a resource of continuum emission data in those directions in which we are interested to determine whether there is a phenomenon in interstellar space, possibly related to the action of the CIV effect, capable of producing continuum radiation.

Our initial search for continuum emission associated with HI structure was limited to an area of sky bounded by l=60° and 180°, b=30° and 70°, which encompasses the bulk of the high-velocity HI complexes and intermediate-velocity gas masses and which also shows an almost complete removal of low-velocity HI produced by whatever has severely disturbed the HI so as to create the high- and intermediate-velocity HI features in the first place.

The study began by searching for associations in position between peaks in the continuum data and HI "clouds," defined as peaks in maps of HI emission at a certain velocity, or averaged over a range of velocities plotted as a function of galactic longitude and latitude (*l-b* maps). It may be expected that comparison of two large data sets composed of many small-scale peaks (sometimes called "clouds" in the case of the HI) both randomly distributed on the sky will always lead to associations being found. However, the galactic foreground HI structure is not random but filamentary, as can be seen in HI area maps presented in reference [7]. Also, very few direct associations between the continuum peaks and HI peaks were noted. Instead, peaks in the continuum data are found to be offset from associated HI peaks by a small angle of order a degree or so, as will be shown in the examples below. Here, we will present three dramatic examples of the relationship between HI structure and continuum emission peaks taken form a larger samples of data [28].

A. A Twisted Filament.

Figure 4 shows a map of the HI emission brightness averaged over the velocity range –100 to –80 km/s as an inverted gray-scale image with a contour map representing the continuum emission brightness overlain. The peak brightness in the HI gray-scale image is of order 5 K km/s. This HI structure is part of a very long twisted filamentary feature extending from $l,b =120°, 62°$ to $175, 67°$, clearly evident in the HI survey data in the velocity range $-100$ to $-80$ km/s [7].

Figure 4 shows that the continuum contours curve around the HI peaks at the left and center while an elongated continuum structure bridges the gap between the peaks at the center and right. These HI peaks are, themselves, linked by a ridge of HI emission visible just to the south of the continuum ridge. The continuum contours appear to thread their way around and through the HI structures. Over the larger area encompassed by this filament for which we have examined detailed maps (from $l=110°$ to $180°$, $b= 60°$ to $70°$), every continuum peak is found to have a closely spaced "companion" HI column density peak at some velocity between $-120$ and $0$ km/s. Their angular relationship is consistent with our hypothesis that continuum emission is produced at the surface of HI gas masses moving with respect to a surrounding medium.

B. HI Structure Around $l,b = 119°, 57°$

The area around $l,b = 119°, 57°$, Fig. 5a, shows a ridge of continuum emission located precisely in a valley between two fragments of high-velocity gas. The HI data represent the brightness temperature at $-121$ km/s. The center velocities of the two HI features are $-127$ and $-118$ km/s while the continuum ridge fits perfectly into the striking minimum between them. It appears that two HI features at different velocities are associated with a continuum structure offset in position with respect to both of them.

A closer look at this area shows that the HI profile structure at low velocities (from 0 to $-20$ km/s) is very complex in the immediate area of the continuum ridge seen in Fig. 5a. This is highlighted by the dramatic presence of a narrow HI emission line at $-8$ km/s at $l,b = 120°, 57°$. Its angular extent has been mapped and is shown as contours in Fig. 5b. Also indicated in that figure is the location of peaks in the HI emission at $-17$ km/s (inverted gray-scale), which appear to be closely tied to the presence of both the high-velocity features and the continuum ridge. The HI feature at $-8$ km/s is unresolved in latitude and slightly elongated in longitude along the continuum ridge line and it is located between two continuum peaks. This close relationship between HI structure at low velocity with that at high-velocity implied by comparison of Figs. 5a and 5b was discovered because the apparent association between the continuum and HI structure drew attention to this region.

C. A Direct Association of Peaks

A striking alignment in position of continuum and HI peaks is shown in Figure 6. Here an HI "cloud" at a velocity of -88 km/s is located at precisely the position of a peak in the continuum data and it has a similar angular extent. This could be regarded as a coincidence but an identical coincidence between the continuum peak and a high-velocity HI structure at –118 km/s is also found. This is shown in the right-hand frame of Fig. 6. No HI emission (<0.2K) is found at velocities between these two features, at –100 km/s.

Here the continuum emission is likely related to the interaction between two HI gas masses along the line of sight and where the CIV effect may have produced enhanced electron densities at the interface. In that case, the continuum emission peak is expected to be coincident in position with the HI gas as seen projected on the sky. In general this would also apply to cases where a moving HI mass is moving along the line-of-sight with respect to the ambient plasma.

VIII. Discussion

The associations between the HI and continuum features shown here are sufficiently tantalizing to merit further attention [28]. We have found that when maps for the high-frequency continuum data (the WMAP ILC signals) and HI over the full study area are compared, the HI structure at virtually all velocities usually avoids peaks in the continuum data. This avoidance is more correctly described as an association between offset structures, expected if the continuum radiation is produced by the CIV effect occurring at the interface between rapidly moving HI and the surrounding plasma. In general the peak continuum radiation would appear offset in angle from the HI feature unless the component of motion of the HI is along the line-of-sight. Note that spatial offsets between HI and parallel HI, dust and Hα filaments have been reported [29].

Unfortunately, there is no unambiguous way to determine the distance to the HI structures and associated continuum peaks discussed above. However, in light of the low velocity of the HI at high galactic latitudes, they are almost certainly local, which means within a few hundred parsecs of the sun.

IX. Conclusions

Based on the data examined so far, several key results have emerged from the present study:

• The 34 km/s wide component in HI profiles appears to be pervasive in the areas of sky at high northern galactic latitudes considered here and in reference [2].

• This component should be identified and removed from the observed HI profile before further Gaussian analysis of HI profiles is carried out.

• After this is done in directions of low total HI column density, the existence of other families of linewidths, with values around 13 and 6 km/s, becomes even more striking than is the case before removal of the 34 km/s component.

• The critical ionization velocity for helium is 34.4 km/s, which strongly suggests the existence of a mechanism in interstellar space in which the CIV effect influences the motion of neutral hydrogen atoms.

• The critical ionization velocity of molecular hydrogen is 38.5 km/s and it, too, may play a role in affecting HI emission profile structure.

• The pervasive presence of the 34 km/s wide component argues for the widespread existence of helium and/or molecular hydrogen in interstellar and nearby intergalactic space, especially in areas where HI gas is moving rapidly with respect to its surroundings.

• The physical cause of the numerical association of these HI emission profile component linewidths with the Bands in the critical ionization velocities remains a mystery.

• Confirmation of the role of the CIV effect in interstellar space is suggested by the discovery of an association between high-frequency continuum emission peaks and HI structure.

• Use of the continuum emission data as pointers has drawn attention to unexpected properties of the interstellar HI that turned out to be uniquely interesting. In future, the continuum data may be used to focus on other interesting areas in interstellar space.

• The high-frequency radio frequency continuum emission data were obtained by the Wilkinson Microwave Anisotropy Probe (WMAP) whose purpose was to study structure in the early universe. If the continuum emission peaks are in fact cosmological in origin, there should be absolutely no relationship between those signals and galactic HI structure. Instead, in the examples shown above, close associations between the two classes of structure are found.

• If confirmed, this argues that the WMAP observations have a strong if not dominant component as a result of processes occurring in interstellar space.

A more extended analysis of the relationship between the WMAP high-frequency continuum emission data and HI structure will be reported in reference [28]

Acknowledgements

I am grateful for the cooperation of Gary Hinshaw and Wayne Landsman of the WMAP team for providing the radio-frequency continuum data in digital form. I also greatly appreciate the encouragement given to me during this work by Joan Schmelz and Butler Burton, and for feedback offered by Tony Peratt, Dave Hogg, Ed Fomalont, Mort Roberts and Mark Ospeck.

Captions:

Figure 1a. Linewidth histogram for northern hemisphere profiles whose center velocities lie between –7 and –30 km/s [2]. The solid curve overlying the bars represent the linewidths found for a sample of compact high-velocity clouds, see text.

Figure 1b. Linewidth histogram for southern hemisphere profiles for components with center velocities between –9 and +3 km/s from [2]. The solid curve overlying the bars represent the linewidths found for a sample of compact high-velocity clouds, see text.

Figure 2. The average HI brightness temperature over a velocity range –10.0 ± 2.5 km/s for an area where the pervasive HI emission profile of width 34 km/s was most readily identified (see text).

Figure 3. Linewidth histogram for low-velocity Gaussian components fit to the HI emission profiles in two areas listed in the text after subtraction of a 34 km/s wide component from each profile.

Figure 4. A section of a twisted HI filament using an inverted gray scale representation of the integrated brightness of the HI emission in the velocity range –100 to –80 km/s. The peak is of order 5 K.km/s. The contours represent the continuum signal levels starting at 0.03 mK with intervals of 0.02 mK.

Figure 5. (a) Inverted gray scale representation of the HI brightness temperature at –121 km/s w.r.t. the l.s.r. with continuum contours at positive values from 0.02 K in intervals of 0.03 K. The two HI peaks are part of high-velocity Complex C. The northern HI peak occurs at –127 km/s while the southern segment peaks at –118 km/s, both with peak brightness temperatures of order 1.2 K, see Fig. 4.

(b) The same continuum contours overlain on an inverted gray-scale image of the column density (peak value ~ $10^{19}$ cm$^{-2}$) of an HI Gaussian component at –17 km/s. The tight set of contours on the ridge of the continuum data indicate the location of an HI component at –8 km/s plotted as a column density of the component whose peak value is ~ 8 x $10^{18}$ cm$^{-2}$. It is unresolved in latitude. At the top right more emission at –8 km/s is found.

Figure 6. Inverted grayscale images of the average HI brightness over a 5 km/s wide range centered approximately at the velocities shown with the continuum contours (from

+0.01 K and increasing in intervals of 0.02 K) overlain. The two HI features, peak brightness of order 4 and 3 K km/s at –118 and –88 km/s, are at very different velocities and yet have the same angular extent and lie precisely on the continuum map peak.

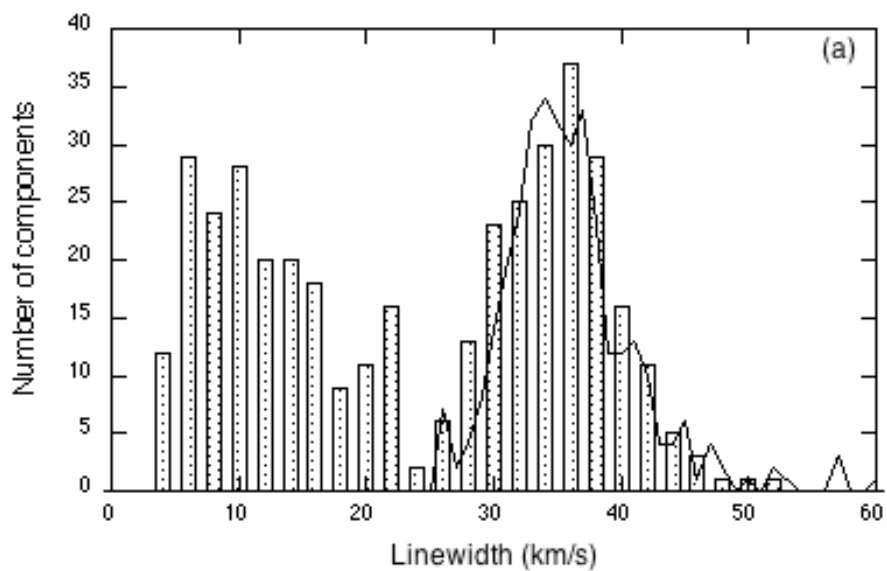

Figure 1a

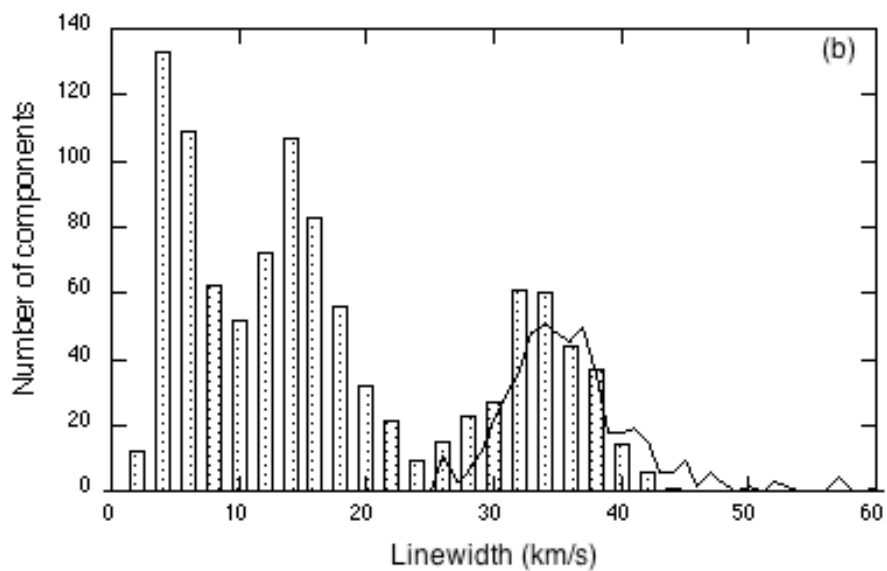

Figure 1b

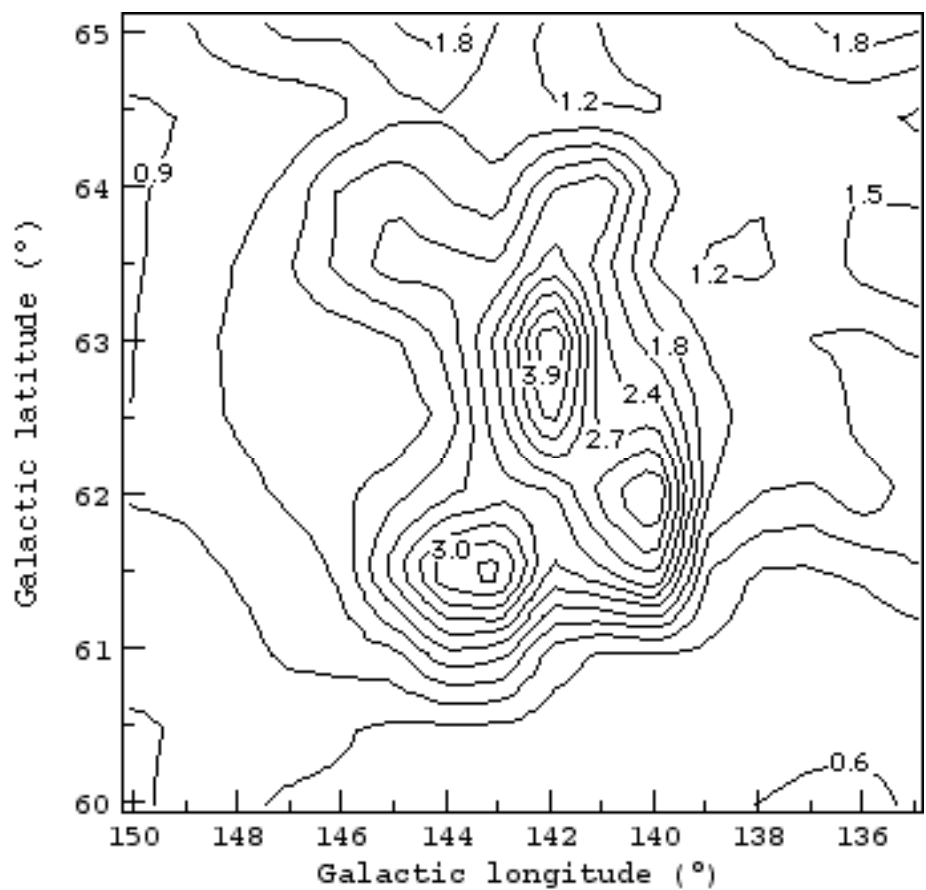

Figure 2

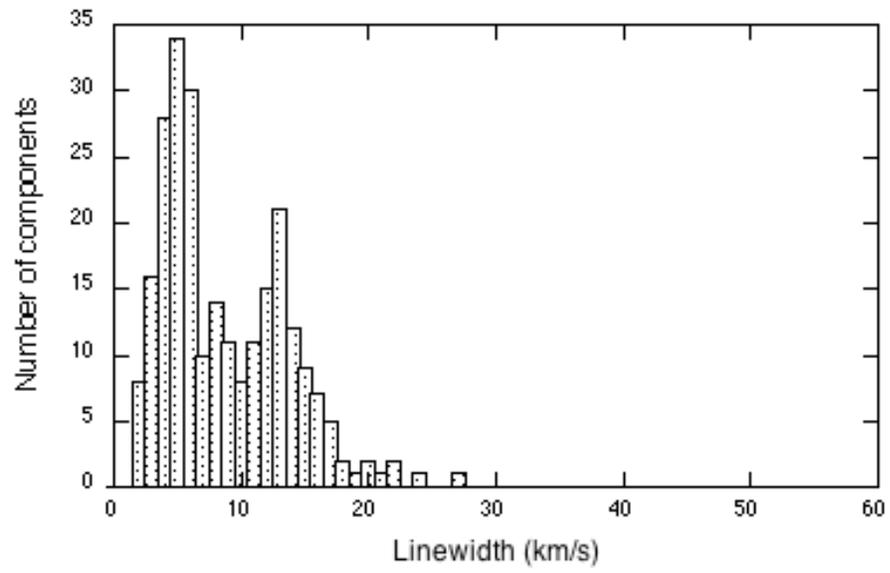

Figure 3

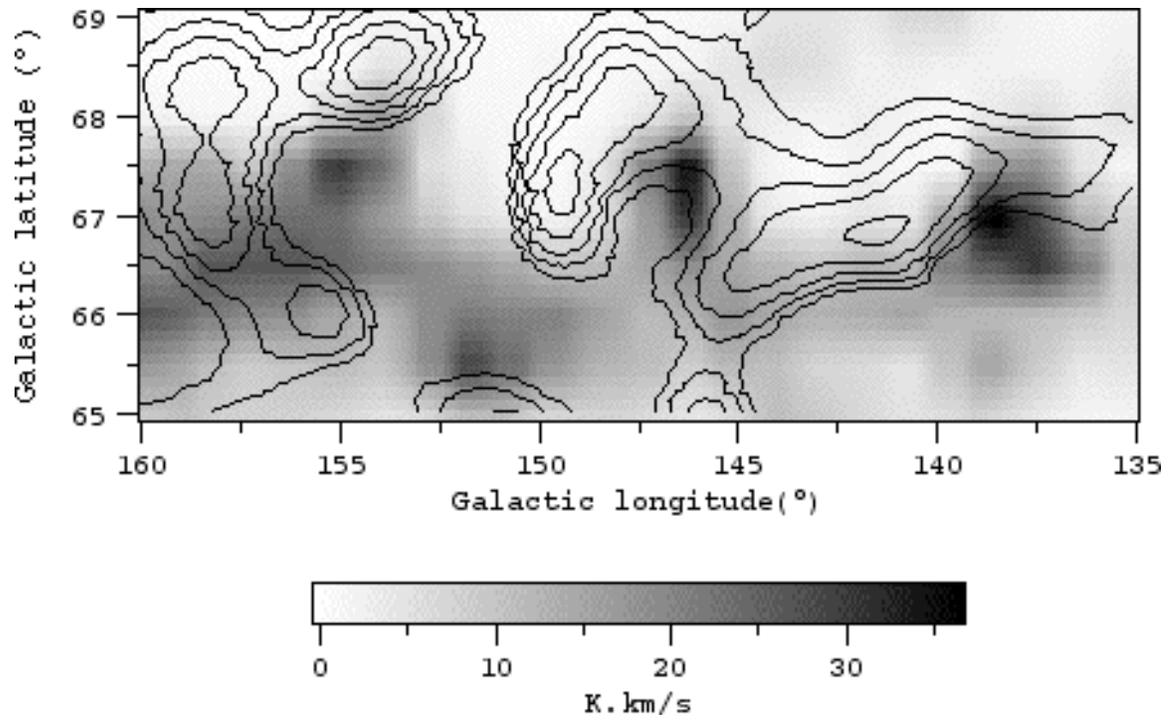

Figure 4

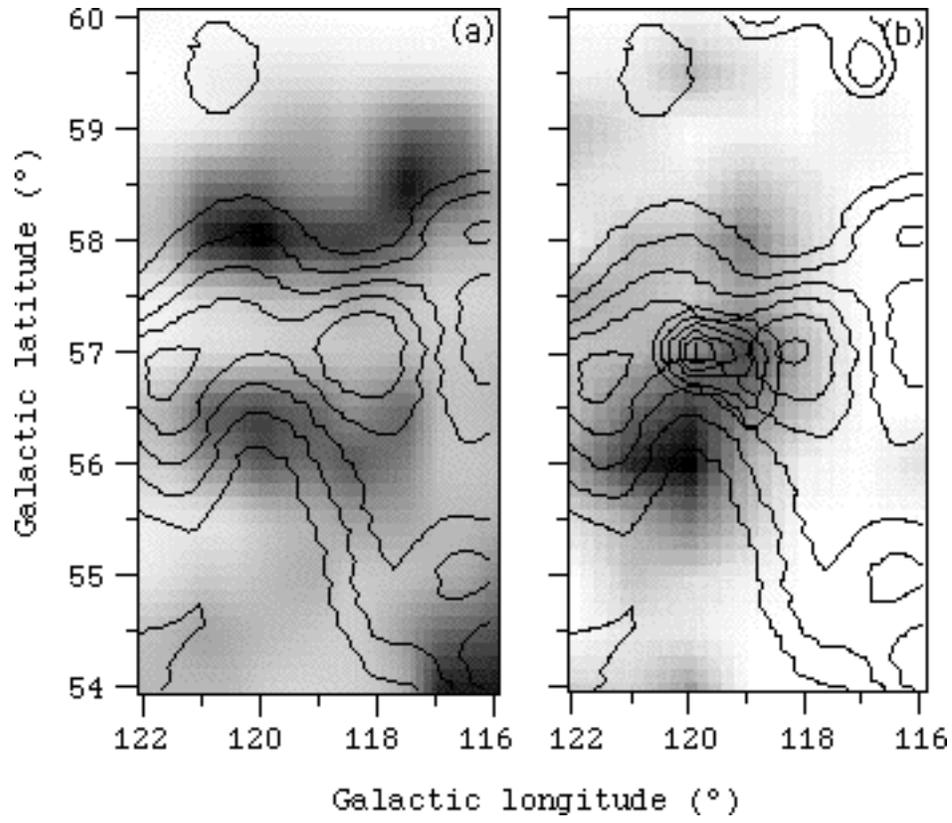

Figure 5

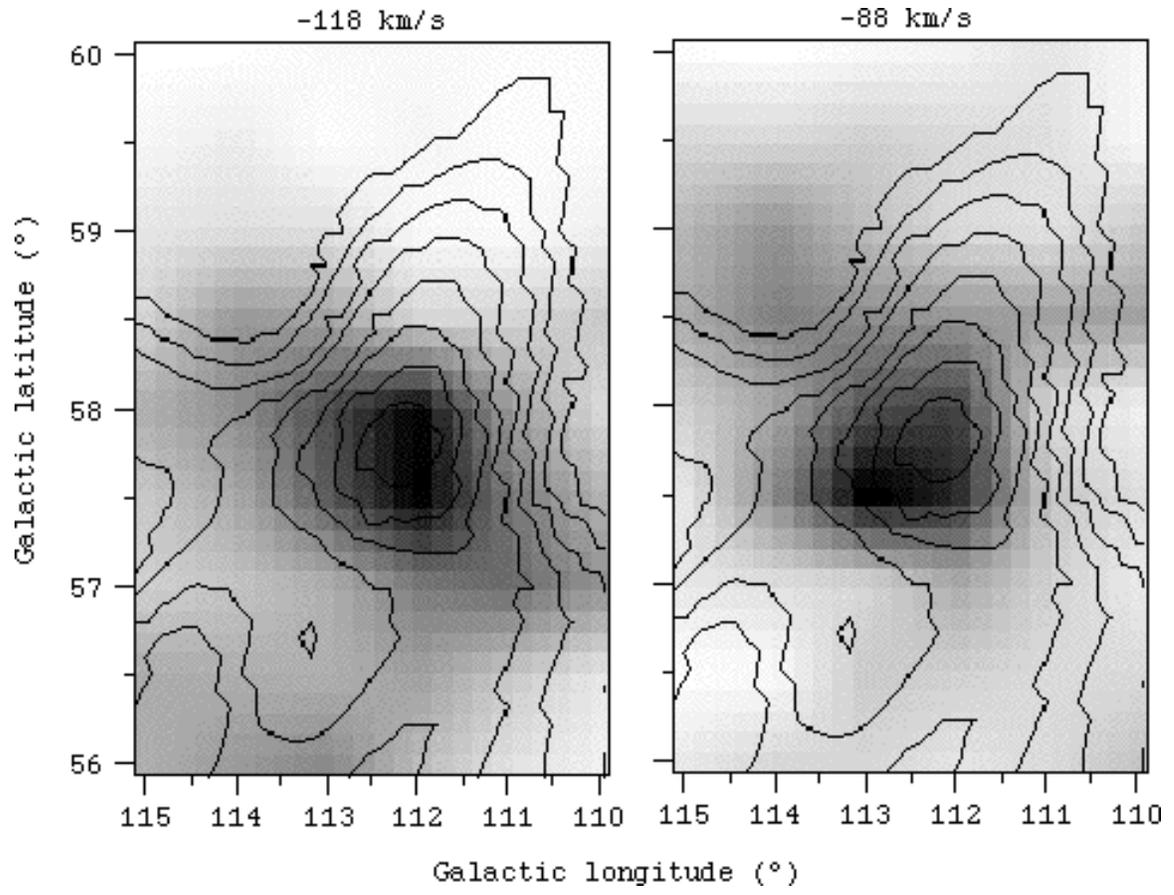

Figure 6